\documentclass[aps,twocolumn,prb]{revtex4}

\usepackage{graphicx}

\begin{document}

\title{Invariant imbedding theory of mode conversion in inhomogeneous
plasmas. II. Mode conversion in cold, magnetized plasmas with
perpendicular inhomogeneity}

\author{Kihong Kim}
\email{khkim@ajou.ac.kr} \affiliation{Department of Molecular
Science and Technology, Ajou University, Suwon, Korea}
\author{Dong-Hun Lee}
\affiliation{Department of Astronomy and Space Science, Kyung Hee
University, Yongin, Korea}

\begin{abstract}
A new version of the invariant imbedding theory for the propagation
of coupled waves in inhomogeneous media is applied to the mode
conversion of high frequency electromagnetic waves into
electrostatic modes in cold, magnetized and stratified plasmas. The
cases where the external magnetic field is applied perpendicularly
to the direction of inhomogeneity and the electron density profile
is linear are considered. Extensive and numerically exact results
for the mode conversion coefficients, the reflectances and the wave
electric and magnetic field profiles inside the inhomogeneous plasma
are obtained. The dependences of mode conversion phenomena on the
magnitude of the external magnetic field, the incident angle and the
wave frequency are explored in detail. 
\end{abstract}

\maketitle

\section{Introduction}

Mode conversion from one type of wave mode to another type of mode
at resonance points in an inhomogeneous plasma is a very important
phenomenon in various branches of plasma
physics.\cite{swanson,budden,ginz,mjol} In a previous
paper,\cite{kim4} we embarked on a project of developing an {\it
exact} theory of mode conversion in inhomogeneous plasmas, which is
based on a new version of the invariant imbedding theory developed
by us earlier.\cite{kim2,kim3} In that paper, we considered the mode
conversion of the simplest kind, that is the conversion of
obliquely-incident $p$-polarized electromagnetic waves into
electrostatic plasma oscillations in cold, unmagnetized and
stratified plasmas.\cite{piliya,fors,hink1,hink3} Using the
invariant imbedding method,\cite{bell,kly0,rammal,kim1,lee,lee2} we
were able to calculate the mode conversion coefficient and the
electromagnetic field distribution for linear and parabolic plasma
density profiles in a numerically exact manner. In some cases, there
were substantial discrepancies between our results and previous
theoretical calculations.

The mode conversion phenomena in cold, unmagnetized plasmas are
described by a single wave equation. A large number of wave
propagation problems in more general situations, however, are
described by two or more coupled wave equations. For the study of
those problems, the invariant imbedding method needs to be
generalized. This generalization was already made in Ref.~7 and has
been successfully applied to various problems where several coupled
waves propagate in inhomogeneous media. The propagation of
electromagnetic waves in magnetized plasmas also involves several
coupled waves. In this paper, we apply the theory developed in
Ref.~7 to the mode conversion of high frequency electromagnetic
waves into electrostatic modes in cold, magnetized and stratified
plasmas.

Even in the relatively simple cases where the plasma is cold and
stratified, the mode conversion of electromagnetic waves in
magnetized plasmas is a highly complex problem. There are three
independent directions in three-dimensional space, which are the
directions of inhomogeneity, the external magnetic field and the
incident wave vector. The wave propagation and mode conversion
phenomena depend on the relative orientations of these three
directions. In the present work, we limit our attention to one
special configuration where the external magnetic field is applied
perpendicularly to the direction of inhomogeneity and plane waves
propagate in the plane spanned by the magnetic field and the
inhomogeneity direction. Other configurations will be considered in
forthcoming papers.

The theory developed in this paper can be applied easily to
multi-species plasmas, which consist of electrons and different
kinds of heavy ions. In this paper, we will consider only the
propagation of high frequency waves and assume heavy ions to be
immobile. Rather complicated, but very interesting mode conversion
phenomena associated with low frequency waves and ion density
oscillations will be subjects of our future publications.

Unlike in the unmagnetized case, the wave propagation in magnetized
plasmas is anisotropic. In a uniform plasma where the wave vector is
not precisely parallel to the external magnetic field, the ordinary
(O) wave and the extraordinary (X) wave are two independent
eigenmodes. When the wave frequency $\omega$ is bigger than the
electron cyclotron frequency $\omega_c$, there can appear the O and
X mode cutoffs and the X mode resonance inside the plasma. On the
other hand, if the wave frequency is smaller than $\omega_c$, the O
mode resonance can appear as well as the O and X mode cutoffs. In
inhomogeneous plasmas, the O and X modes are no longer true
eigenmodes and are coupled to each other. These modes can also be
converted to upper hybrid oscillations at the resonances. The
presence of both a strong coupling between different modes and the
singularity associated with resonance points makes the solution of
the mode conversion problem very difficult. As will be demonstrated
later, our theory allows us to obtain an exact solution of this
difficult problem in a remarkably simple and efficient manner.

\begin{figure}
\includegraphics[width=8cm]{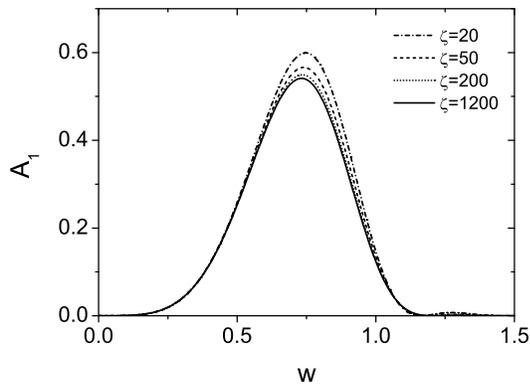}
\caption{Mode conversion coefficient $A_1$ for normally-incident $s$
waves as a function of the parameter $w= \zeta^{1/3}\sqrt{Y}$ for
several values of $\zeta= \omega \Lambda/c$ and for
$\tilde{\nu}=10^{-8}$ and $L/\Lambda=20$. $A_1$ approaches a
universal curve as $\zeta$ becomes large.} \label{f.1}
\end{figure}

Mode conversion in magnetized plasmas has been studied extensively
over several
decades.\cite{piliya2,woo,weitzner,kitagawa,window1,window2,wong,
maggs,mjol3,mjol4,mjol2,seliga,hansen,johnson0,antani,ueda,yin,
golda,kskim} We will not attempt to give a detailed review of the
literature here. Many of the previous works have used one
approximation scheme or another, most frequently the WKB method. In
a series of papers, Mj{\o}lhus calculated the mode conversion
coefficient for cold, magnetized plasmas primarily using the WKB
method.\cite{mjol3,mjol4,mjol2} This work, which was reviewed in
Ref.~4 together with theoretical results of many other researchers,
often predicts a complete mode conversion for certain parameters, in
great contrast with our results. This discrepancy might be explained
in large part by one crucial difference between Mj{\o}lhus' and our
calculations. In the present paper, we consider only the cases where
linearly-polarized waves are incident from a vacuum onto a
stratified plasma, whereas Mj{\o}lhus and other researchers have
considered the cases where O or X waves are incident from a uniform
plasma. Upon entering the plasma, linearly-polarized waves will be
split into the O and X mode components. In the case of perpendicular
inhomogeneity considered in this paper and when the waves are
incident on a stratified plasma normally, mode conversion is caused
solely by the X mode component. In this case, we find a perfect
agreement between our result for incident $s$ waves shown in
Fig.~\ref{f.1} and Mj{\o}lhus' result for incident X waves. When the
waves are incident obliquely and for sufficiently large values of
the parameter $w$ defined in Sec.~\ref{sec4a}, mode conversion is
dominated by the incident O mode. This case is closely related to
the theory of radio windows in the ionosphere and
magnetosphere.\cite{budden,ginz,mjol,window1,window2} We find that
the critical parameter values where the mode conversion coefficient
takes the maximum value agrees extremely well with the analytical
formula, Eq.~(\ref{eq:win3}), obtained by previous theories, for
sufficiently large $w$. This is another evidence for the consistency
between our theory and previous theories. For quantitative
comparisons with previous results, we need to generalize our method
to the cases where O or X waves are incident from a uniform plasma.
This generalization will be pursued in future research. Since our
theory is capable of producing exact results for realistic plasma
density profiles, we expect it will be useful in the interpretation
of various radio modification experiments.

In Sec.~\ref{sec2}, we present the details of a new version of the
invariant imbedding theory for the propagation of coupled waves in
stratified media. In Sec.~\ref{sec3a}, we introduce the coupled wave
equations for electromagnetic waves in cold, magnetized plasmas with
perpendicular inhomogeneity. In Sec.~\ref{sec3b}, we describe two
linear plasma density profiles used for the calculations in this
paper. The results of our calculations are presented in
Sec.~\ref{sec4}. In Sec.~\ref{sec4a}, we study the magnetic field
dependence of mode conversion coefficients. In Sec.~\ref{sec4b}, we
consider the dependence of mode conversion coefficients on the
incident angle in detail. We also consider the spatial dependence of
electric and magnetic field intensities. In Sec.~\ref{sec4c}, we
discuss the frequency dependence of mode conversion coefficients. We
conclude the paper in Sec.~\ref{sec5}. In Appendix \ref{sec6}, we
give a detailed derivation of Eq.~(\ref{eq:matrix}).

\section{Invariant imbedding theory of the propagation of
coupled waves in stratified media} \label{sec2}

In Ref.~7, we have developed a new version of the invariant
imbedding theory for the propagation of coupled waves in
arbitrarily-inhomogeneous stratified media. Since this theory will
play a central role in our development of the theory of mode
conversion, we give a detailed presentation of it here.

We consider $N$ coupled waves propagating in a stratified medium,
where all physical parameters depend on only one spatial coordinate.
We choose this coordinate as the $z$ axis and assume the
inhomogeneous medium of thickness $L$ lies in $0\le z\le L$. We also
assume that all waves propagate in the $xz$ plane. The $x$ component
of the wave vector, $q$, is then a constant. In a large class of
interesting problems, the wave equation of $N$ coupled waves in
stratified media has the form
\begin{eqnarray}
{{d^2 \psi}\over{dz^2}}-\frac{d\cal E}{dz}{\cal E}^{-1}(z)
\frac{d\psi}{dz}+\left[{\cal E}(z)K^2{\cal M}(z)-q^2I\right]\psi=0,
\label{eq:wave}
\end{eqnarray}
where $\psi=(\psi_1,\cdots,\psi_N)^T$ is a vector wave function and
${\cal E}$ and ${\cal M}$ are $N\times N$ matrix functions. We
assume that the waves are incident from a vacuum region where $z>L$
and transmitted to another vacuum region where $z<0$. $I$ is a unit
matrix and $K$ is a diagonal matrix such that
$K_{ij}=k_i\delta_{ij}$, where $k_i$ is the magnitude of the vacuum
wave vector for the $i$-th wave. In the problem considered in this
paper, all $k_i$'s are equal. However, there exist many cases where
$k_i$'s take different values. It is useful to rewrite
Eq.~(\ref{eq:wave}) as
\begin{eqnarray}
\frac{d}{dz}\left[{\cal E}^{-1}(z){{d
\psi}\over{dz}}\right]+\left[K^2{\cal M}(z)-q^2{\cal
E}^{-1}(z)\right]\psi=0. \label{eq:waveb}
\end{eqnarray}
From this form, we can see that ${\cal E}^{-1}d\psi/dz$ is
continuous everywhere except at some singular points.

Eq.~(\ref{eq:wave}) is suitable when the matrices ${\cal E}$ and
${\cal M}$ do not contain the constant $q$. In the cases where the
$q$ dependence cannot be separated as simply as in
Eq.~(\ref{eq:wave}), it is more convenient to write the wave
equation as
\begin{eqnarray}
{{d^2 \psi}\over{dz^2}}-\frac{d\cal E}{dz}{\cal E}^{-1}(z)
\frac{d\psi}{dz}+{\cal E}(z)K^2{\cal M}(z)\psi=0 \label{eq:wave2}
\end{eqnarray}
and define ${\cal E}$ and ${\cal M}$ as $q$-dependent quantities. By
assigning ${\cal E}(z)$ and ${\cal M}(z)$ suitably, we are able to
describe various kinds of waves in a number of stratified media
using either Eq.~(\ref{eq:wave}) or (\ref{eq:wave2}). A wide variety
of mode conversion phenomena observed in space and laboratory
plasmas can be studied using these equations.

We generalize Eqs.~(\ref{eq:wave}) and (\ref{eq:wave2}) slightly, by
replacing the vector wave function $\psi$ by an $N\times N$ matrix
wave function $\Psi$, the $j$-th column vector
$(\Psi_{1j},\cdots,\Psi_{Nj})^T$ of which represents the wave
function when the incident wave consists only of the $j$-th wave of
unit amplitude. Our main interest is in the reflection and
transmission coefficient matrices $r=r(L)$ and $t=t(L)$. The wave
function in the incident and transmissive regions are expressed in
terms of $r$ and $t$:
\begin{eqnarray}
\Psi(z)=\left\{ \begin{array}{ll} e^{iP(L-z)} +e^{iP(z-L)}~r,
&~z>L\\
e^{-iPz}~t, &~z<0 \end{array} \right., \label{eq:psi}
\end{eqnarray}
where $P$ is a diagonal matrix satisfying $P_{ij}=p_i\delta_{ij}$
and $p_i$ is the negative $z$ component of the vacuum wave vector
for the $i$-th wave. It is important to write the products of two
matrices, $e^{iP(z-L)}~r$ and $e^{-iPz}~t$, in the given order. From
Eq.~(\ref{eq:psi}), we obtain $\Psi(L)=I+r$ and $\Psi(0)=t$.

Let us introduce a matrix
\begin{eqnarray}
g(z,z^\prime)=\left\{ \begin{array}{ll} {\mathcal T}\exp\left[i
\int_{z^\prime}^zdz^{\prime\prime} ~{\cal
E}(z^{\prime\prime})P\right],
&~z>z^\prime\\
\tilde{\mathcal T}\exp\left[-i \int_{z^\prime}^zdz^{\prime\prime}
~{\cal E}(z^{\prime\prime})P\right], &~z<z^\prime \end{array}
\right. \label{eq:gf}
\end{eqnarray}
where $\mathcal T$ and $\tilde{\mathcal T}$ are the time-ordering
and anti-time-ordering operators respectively. When applied to a
product of matrices which are functions of $z$, $\mathcal T$
($\tilde{\mathcal T}$) arranges the matrices in the order of
decreasing (increasing) $z$. For instance, $\mathcal T\left[{\cal
E}(z_1){\cal E}(z_2)\right]={\cal E}(z_2){\cal E}(z_1)$ and
$\tilde{\mathcal T}\left[{\cal E}(z_2){\cal E}(z_1)\right]={\cal
E}(z_1){\cal E}(z_2)$, if $z_2>z_1$. Though, strictly speaking,
$\mathcal T$ has to be called a space-ordering operator, we keep
using the standard terminology, time-ordering operator, commonly
used in quantum mechanics.\cite{qm} Expanding the matrix
exponentials in Eq.~(\ref{eq:gf}) in a Taylor series and using the
definitions of $\mathcal T$ and $\tilde{\mathcal T}$, we can easily
prove that $g(z,z^\prime)$ satisfies the differential equations
\begin{eqnarray}
\frac{\partial}{\partial z}g(z,z^\prime)&=&i~ {\rm sgn}(z-z^\prime)
~{\mathcal E}(z)Pg(z,z^\prime),\nonumber\\
\frac{\partial}{\partial z^\prime}g(z,z^\prime)&=& -i ~{\rm
sgn}(z-z^\prime)~ g(z,z^\prime){\mathcal E}(z^\prime)P.
\label{eq:gf2}
\end{eqnarray}

Using Eqs.~(\ref{eq:gf}) and (\ref{eq:gf2}), the wave equation
(\ref{eq:wave}) is transformed exactly to an integral equation
\begin{widetext}
\begin{eqnarray}
&&\Psi(z,L)=g(z,L)\nonumber\\
&&~~-{i\over 2}\int_0^L dz^\prime g(z,z^\prime)\left[ {\cal
E}(z^\prime)P-P{\cal M}(z^\prime)-q^2P^{-1}{\cal M}(z^\prime)
+q^2P^{-1}{\cal E}^{-1}(z^\prime)\right]\Psi(z^\prime,L),
\label{eq:inteq}
\end{eqnarray}
\end{widetext}
where we have considered $\Psi$ as a function of both $z$ and $L$.
We replace $L$ by a variable $l$ and take a derivative of this
equation with respect to $l$. Then we obtain
\begin{eqnarray}
&&\frac{\partial \Psi(z,l)}{\partial
l}=i\Psi(z,l)\alpha(l)+\Phi(z,l), \label{eq:main0}
\end{eqnarray}
where
\begin{widetext}
\begin{eqnarray}
&&\alpha(l)={\cal E}(l)P-{1\over 2}\left[{\cal E}(l)P-P{\cal M}(l)
-q^2P^{-1}{\cal M}(l) +q^2P^{-1}{\cal
E}^{-1}(l)\right]\Psi(l,l),\nonumber\\ &&\Phi(z,l)=-{i\over
2}\int_0^l dz^\prime g(z,z^\prime)\left[ {\cal E}(z^\prime)P-P{\cal
M}(z^\prime)-q^2P^{-1}{\cal M}(z^\prime) +q^2P^{-1}{\cal
E}^{-1}(z^\prime)\right]\Phi(z^\prime,l).
\end{eqnarray}
\end{widetext}
We notice that $\Phi(z,l)$ satisfies an equation similar to
Eq.~(\ref{eq:inteq}) except that there is no source term (that is,
$g(z,l)$). This implies $\Phi(z,l)=0$ and then we have
\begin{equation}
\frac{\partial \Psi(z,l)}{\partial l}=i\Psi(z,l)\alpha(l).
\label{eq:main}
\end{equation}
Taking now the derivative of $\Psi(l,l)$ with respect to $l$, we
obtain
\begin{eqnarray}
\frac{d \Psi(l,l)}{d l}&=&\frac{\partial \Psi(z,l)}{\partial z}
\Bigg\vert_{z=l}+ \frac{\partial \Psi(z,l)}{\partial
l}\Bigg\vert_{z=l}\nonumber\\& =&i{\cal
E}(l)P\left[r(l)-I\right]+i\Psi(l,l)\alpha(l),
\end{eqnarray}
where we have used Eq.~(\ref{eq:main}) and the fact that ${\mathcal
E}^{-1}(z)~\partial\Psi(z,l)/\partial z$ is continuous at $z=l$.
Since $\Psi(l,l)=I+r(l)$, we find the invariant imbedding equation
satisfied by $r(l)$:
\begin{widetext}
\begin{eqnarray}
{{dr}\over{dl}}&=&i\left[r(l){\cal E}(l)P+{\cal E}(l)Pr(l)\right]
\nonumber\\
&&-{i \over 2}[r(l)+I]\left[ {\cal E}(l)P-P{\cal
M}(l)-q^2P^{-1}{\cal M}(l) +q^2P^{-1}{\cal
E}^{-1}(l)\right][r(l)+I]. \label{eq:imbedr}
\end{eqnarray}
\end{widetext}
Similarly by setting $z=0$ in Eq.~(\ref{eq:main}), we find the
invariant imbedding equation for $t(l)$ ($=\Psi(0,l)$):
\begin{widetext}
\begin{eqnarray}
{{dt}\over{dl}}&=&it(l){\cal E}(l)P \nonumber\\&&-{i \over
2}t(l)\left[ {\cal E}(l)P-P{\cal M}(l)-q^2P^{-1}{\cal M}(l)
+q^2P^{-1}{\cal E}^{-1}(l)\right][r(l)+I]. \label{eq:imbedt}
\end{eqnarray}
\end{widetext}
These equations are supplemented with the initial conditions,
$r(0)=0$ and $t(0)=I$.

We integrate the coupled differential equations (\ref{eq:imbedr})
and (\ref{eq:imbedt}) numerically from $l=0$ to $l=L$ using the
initial conditions and obtain the reflection and transmission
coefficient matrices $r$ and $t$ as functions of $L$. The invariant
imbedding method can also be used in calculating the field amplitude
$\Psi(z)$ inside the inhomogeneous medium. Rewriting
Eq.~(\ref{eq:main}), we get
\begin{widetext}
\begin{eqnarray}
{{\partial\Psi(z,l)}\over{\partial l}}&=&i\Psi(z,l){\cal E}(l)P
\nonumber\\&& -{i \over 2}\Psi(z,l)\left[ {\cal E}(l)P-P{\cal
M}(l)-q^2P^{-1}{\cal M}(l) +q^2P^{-1}{\cal
E}^{-1}(l)\right][r(l)+I]. \label{eq:imbedf}
\end{eqnarray}
\end{widetext}
For a given $z$ ($0<z<L$), the field amplitude $\Psi(z,L)$ is
obtained by integrating this equation from $l=z$ to $l=L$ using the
initial condition $\Psi(z,z)=I+r(z)$.

Our invariant imbedding equations,
Eqs.~(\ref{eq:imbedr}-\ref{eq:imbedf}), have been applied
successfully to the study of the propagation of electromagnetic
waves in multilayer systems made of chiral media and of
phase-conjugating media.\cite{kim3}

\section{Coupled wave equations for cold, magnetized plasmas
with perpendicular inhomogeneity}\label{sec3}

\subsection{Coupled wave equations}\label{sec3a}

Eqs.~(\ref{eq:imbedr}), (\ref{eq:imbedt}) and (\ref{eq:imbedf}) are
the starting point in our exact analysis of a variety of wave
coupling and mode conversion phenomena. In the rest of this paper,
we apply these equations to the high frequency wave propagation and
mode conversion in cold, magnetized plasmas. For the sake of
simplicity, we restrict our attention to the cases where the
external magnetic field is applied perpendicularly to the direction
of inhomogeneity.

We assume that the plasma density varies only in the $z$ direction
and the uniform external magnetic field ${\bf B}_0$ ($=B_0\hat{\bf
x}$) is directed in the $x$ direction. The cold plasma dielectric
tensor, $\epsilon$, for high frequency waves in the present geometry
is written as
\begin{equation}
\epsilon=\pmatrix{\epsilon_3 &0&0\cr 0&\epsilon_1& i\epsilon_2\cr 0&
-i\epsilon_2&\epsilon_1}, \label{eq:dtensor}
\end{equation}
where
\begin{eqnarray}
\epsilon_1&=&1-\frac{\omega_p^2\left(\omega+i\nu\right)}
{\omega\left[\left(\omega+i\nu\right)^2-\omega_c^2\right]},\nonumber\\
\epsilon_2&=&\frac{\omega_p^2\omega_c}
{\omega\left[\left(\omega+i\nu\right)^2-\omega_c^2\right]},\nonumber\\
\epsilon_3&=&1-\frac{\omega_p^2}{\omega\left(\omega+i\nu\right)}.
\end{eqnarray}
We note that the indices 1, 2 and 3 in $\epsilon_1$, $\epsilon_2$
and $\epsilon_3$ have no direct relationship with the spatial
coordinates $x$, $y$ and $z$. The constant $\nu$ is the
phenomenological collision frequency. In Gaussian units, the
electron plasma frequency $\omega_p$ and the electron cyclotron
frequency $\omega_c$ are given by
\begin{equation}
\omega_p^2=\frac{4\pi e^2}{m_e}n(z),~~~\omega_c=\frac{eB_0}{m_e c},
\end{equation}
where $m_e$ and $-e$ are the electron mass and charge respectively.
The spatial inhomogeneity of plasmas enters through the $z$
dependence of the electron number density $n$.

For monochromatic waves of frequency $\omega$, the wave equations
satisfied by the electric and magnetic fields in cold, magnetized
plasmas have the form
\begin{eqnarray}
&&-\nabla\times\left(\nabla\times{\bf E}\right)+{\omega^2\over
c^2}\epsilon\cdot{\bf E}=0,\nonumber\\
&& -\nabla\times\left(\epsilon^{-1}\cdot\nabla\times{\bf
B}\right)+{\omega^2\over c^2}{\bf B}=0.\label{eq:cwe}
\end{eqnarray}
In this paper, we restrict our interest to the cases where plane
waves propagate parallel to the $xz$ plane, which is sometimes
called the magnetic meridian plane. We assume that all field
components depend on $x$ and $t$ through a factor
$e^{i[(k\sin\theta)x-\omega t]}$, where $k=\omega/c$ and $\theta$ is
the incident angle, and have no $y$ dependence. In this situation,
we can eliminate $E_x$, $E_z$, $B_x$ and $B_z$ from
Eq.~(\ref{eq:cwe}) and obtain two coupled wave equations satisfied
by the $z$-dependent field amplitudes $E_y=E_y(z)$ and $B_y=B_y(z)$,
which turn out to have precisely the same form as
Eq.~(\ref{eq:wave2}) with
\begin{eqnarray}
&&\psi=\pmatrix{ E_y \cr B_y \cr}, ~~
K=\frac{\omega}{c}I,\nonumber\\
&&{\cal E}=\pmatrix{1&0\cr 0&\epsilon_3\cr},\nonumber\\
&&{\cal
M}=\pmatrix{\epsilon_1-\frac{{\epsilon_2}^2}{\epsilon_1}-\sin^2\theta
& -i{\epsilon_2\over\epsilon_1}\sin\theta \cr
i{\epsilon_2\over\epsilon_1}\sin\theta &
1-\frac{\sin^2\theta}{\epsilon_1}\cr}. \label{eq:matrix}
\end{eqnarray}
A derivation of this crucial result is given in Appendix A.

We use Eqs.~(\ref{eq:imbedr}), (\ref{eq:imbedt}) and
(\ref{eq:matrix}) in calculating the reflection and transmission
coefficients. In our notation, $r_{11}$ ($r_{21}$) is the reflection
coefficient when the incident wave is $E_y$ (that is, $s$-polarized)
and the reflected wave is $E_y$ ($B_y$). Similarly, $r_{22}$
($r_{12}$) is the reflection coefficient when the incident wave is
$B_y$ (that is, $p$-polarized) and the reflected wave is $B_y$
($E_y$). Similar definitions are applied to the transmission
coefficients. The reflectances and transmittances are defined by
$R_{ij}=\vert r_{ij}\vert^2$ and $T_{ij}=\vert t_{ij}\vert^2$. When
the dielectric permittivities of the incident region and the
transmissive region are the same, we can calculate the wave
absorption by $A_j\equiv 1-R_{1j}-R_{2j}-T_{1j}-T_{2j}$ ($j=1,2$).
If a mode conversion occurs, this quantity is nonzero even in the
limit where the damping constant $\nu$ goes to zero.\cite{kim4} In
that limit, we will call $A_j$ as the mode conversion coefficient.

Finally, we use Eqs.~(\ref{eq:imbedf}) and (\ref{eq:matrix}) in
calculating the electromagnetic field amplitudes $E_y(z)$ and
$B_y(z)$ inside the inhomogeneous region. Once we obtain $E_y$ and
$B_y$, we can use Eq.~(\ref{eq:xz}) to calculate other field
components $E_x$, $E_z$, $B_x$ and $B_z$.

\subsection{Plasma density profiles}\label{sec3b}

Our theory can be applied to any function $n(z)$ for the electron
density profile. In the present work, we confine our calculations to
two kinds of linear profiles. In the first one, we assume that the
density is given by
\begin{equation}
n(z)=\left\{\begin{array}{ll}
0 & ~~\mbox{if $z>L$}\\
n_0\frac{L-z}{\Lambda}
& ~~\mbox{if $0 \le z \le L$}\\
0 & ~~\mbox{if $z<0$}\end{array}\right., \label{eq:profile1}
\end{equation}
where the constant $\Lambda$ is the scale length for the linear
density profile, and the wave frequency is fixed at the value of the
local plasma frequency at $z=L-\Lambda$, that is $\omega=(4\pi n_0
e^2/m_e)^{1/2}$. In this case, the dielectric tensor for $0\le z\le
L$ is given by Eq.~(\ref{eq:dtensor}) with
\begin{eqnarray}
\epsilon_1&=&1-\frac{L-z}{\Lambda}\frac{1+i\tilde{\nu}}
{\left(1+i\tilde{\nu}\right)^2-Y^2},\nonumber\\
\epsilon_2&=&\frac{L-z}{\Lambda}\frac{Y}
{\left(1+i\tilde{\nu}\right)^2-Y^2},\nonumber\\
\epsilon_3&=&1-\frac{L-z}{\Lambda}\frac{1}{1+i\tilde{\nu}},
\end{eqnarray}
where $\tilde\nu\equiv \nu/\omega$ and $Y\equiv\omega_c/\omega$. For
$z>L$ or $z<0$, the dielectric tensor is equal to a unit matrix. We
will use this tensor in studying the dependencies of various
quantities on the incident angle and the strength of the external
magnetic field.

We are also interested in the frequency dependence of mode
conversion phenomena. For this purpose, we consider a linear density
profile for $0\le z\le L$ such that the plasma frequency satisfies
\begin{eqnarray}
\omega_p= 2\pi f_0
\left[1+24\left(1-\frac{z}{L}\right)\right]^{1/2},
\label{eq:profile2}
\end{eqnarray}
where $f_0$ is the local (nonangular) plasma frequency at $z=L$
measured in MHz. We introduce nonangular frequencies
\begin{equation}
f\equiv\frac{\omega}{2\pi},~~ f_c\equiv\frac{\omega_c}{2\pi},~~
f_\nu\equiv\frac{\omega_\nu}{2\pi},
\end{equation}
also measured in MHz.
Then, for $0\le z\le L$, the dielectric tensor is given by
Eq.~(\ref{eq:dtensor}) with
\begin{eqnarray}
\epsilon_1&=&1- \frac{f_0^2\left(f+if_\nu\right)}
{f\left[\left(f+if_\nu\right)^2-f_c^2\right]}
\left[1+24\left(1-\frac{z}{L}\right)\right],\nonumber\\
\epsilon_2&=&\frac{f_0^2f_c}
{f\left[\left(f+if_\nu\right)^2-f_c^2\right]}
\left[1+24\left(1-\frac{z}{L}\right)\right],\nonumber\\
\epsilon_3&=&1- \frac{f_0^2}{f\left(f+if_\nu\right)}
\left[1+24\left(1-\frac{z}{L}\right)\right].
\end{eqnarray}
If $z>L$ or $z<0$, the dielectric tensor is equal to a unit matrix.
For specific calculations, we will assume $f_0=1$ MHz and $f_c=1.5$
MHz. Then the local plasma frequency varies from 1 MHz at $z=L$ to 5
MHz at $z=0$.

\section{Results}\label{sec4}

\subsection{Magnetic field dependence}\label{sec4a}

In this subsection, we study the dependence of mode conversion
coefficients on the magnitude of the external magnetic field $B_0$,
or equivalently on the cyclotron frequency $\omega_c$. We first
consider the simplest case where linearly-polarized plane waves are
incident normally (i.e. $\theta=0$) on a layer of inhomogeneous
plasma with the linear density profile given by
Eq.~(\ref{eq:profile1}). In this case, it is well-known that there
is a resonance associated with X waves if the wave frequency is
equal to the upper hybrid frequency
$\omega_{uh}=(\omega_p^2+\omega_c^2)^{1/2}$, whereas there is no
resonance associated with O waves. For $L>\Lambda$, the resonance
condition $\omega=\omega_{uh}$ can be satisfied if $Y<1$ (i.e.
$\omega>\omega_c$). The coordinate of the resonance point $z_R$ is
given by
\begin{equation}
z_R=L-\Lambda\left(1-Y^2\right). \label{eq:respt}
\end{equation}
Since we are considering the cases where waves are incident from a
vacuum region, we assume that the incident wave is either $s$- or
$p$-polarized. In the case of normal incidence, $p$ waves are the
same as O waves and $s$ waves are similar to X waves. Therefore we
expect that there is no mode conversion when $p$ waves are incident
normally, whereas some portion of normally-incident $s$ waves will
convert to upper hybrid oscillations.

Actual calculations using our invariant imbedding theory show that
the $p$ wave absorption $A_2$ vanishes indeed when the damping
constant $\tilde \nu$ is very small. In Fig.~\ref{f.1}, we plot the
$s$ wave mode conversion coefficient $A_1$ as a function of the
dimensionless parameter $w\equiv \zeta^{1/3}\sqrt{Y}$ ($\propto
\sqrt{B_0}$) for several values of $\zeta\equiv \omega \Lambda/c$
and for $\theta=0$, $\tilde{\nu}=10^{-8}$ and $L/\Lambda=20$. We
notice that $A_1$ approaches a {\it universal} curve as $\zeta$
increases to large values. We find this universal curve agrees
remarkably well with Fig.~4 presented in Ref.~4.

\begin{figure}
\includegraphics[width=8cm]{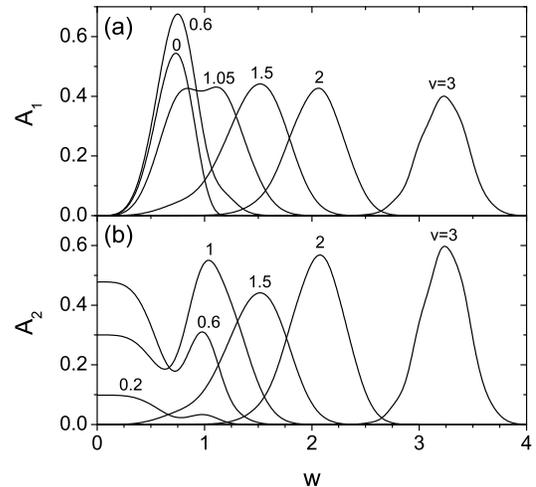}
\caption{Mode conversion coefficient $A_1$ ($A_2$) for
obliquely-incident $s$ ($p$) waves as a function of $w$ for several
values of $v= \zeta^{1/3}\sin\theta$ and for $\zeta=500$,
$\tilde{\nu}=10^{-8}$ and $L/\Lambda=20$.} \label{f.2}
\end{figure}

Next we consider the cases where linearly-polarized waves are
incident obliquely at an angle $\theta$. In those cases, both $s$
and $p$ waves can generate X mode components inside the
inhomogeneous plasma and be converted to upper hybrid oscillations
at $z=z_R$. In Fig.~\ref{f.2}, we show the $s$ ($p$) wave mode
conversion coefficient $A_1$ ($A_2$) as a function of $w$ for
several values of the parameter $v\equiv \zeta^{1/3}\sin\theta$ and
for $\zeta=500$, $\tilde{\nu}=10^{-8}$ and $L/\Lambda=20$. As the
incident angle increases, $A_2$ grows from zero to finite values.

\begin{figure}
\includegraphics[width=8cm]{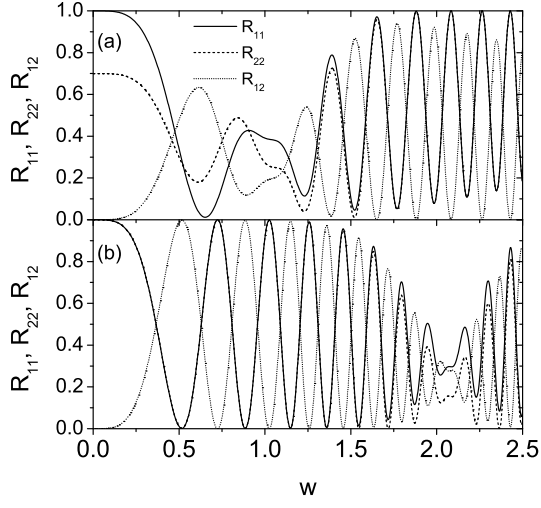}
\caption{Reflectances $R_{11}$, $R_{22}$ and $R_{12}$ ($=R_{21}$) as
a function of $w$ for $\zeta=500$, $\tilde{\nu}=10^{-8}$,
$L/\Lambda=20$ and (a) $v=1$, (b) $v=2$.} \label{f.3}
\end{figure}

In Fig.~\ref{f.3}, we plot the reflectances $R_{11}$, $R_{22}$ and
$R_{12}$ for $\zeta=500$, $\tilde{\nu}=10^{-8}$ and $L/\Lambda=20$.
The parameter $v$ is equal to 1 in Fig.~3(a) and 2 in Fig.~3(b). It
can be proved that $R_{21}$ is always identical to $R_{12}$. We
observe that in the parameter region where there is no mode
conversion, the reflectances are rapidly oscillating functions of
$w$.

\subsection{Incident angle dependence
and the electromagnetic field distribution}\label{sec4b}

\begin{figure}
\includegraphics[width=8cm]{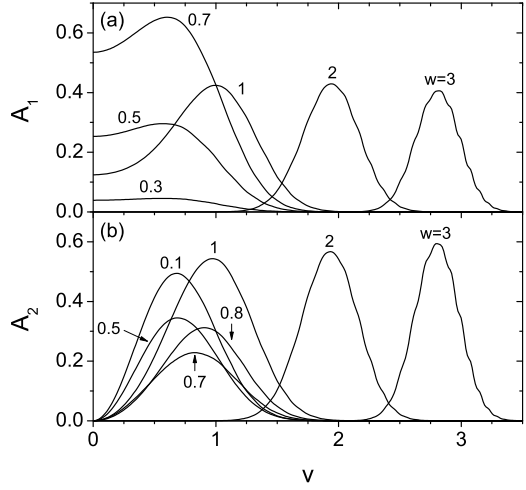}
\caption{Mode conversion coefficients $A_1$ and $A_2$ as a function
of $v$ for several values of $w$ and for $\zeta=500$,
$\tilde{\nu}=10^{-8}$ and $L/\Lambda=20$.} \label{f.4}
\end{figure}

In this subsection, we consider the dependence of mode conversion
coefficients on the incident angle $\theta$ in detail. We also
consider the spatial dependence of electric and magnetic field
intensities associated with the electromagnetic wave. In
Fig.~\ref{f.4}, we show the $s$ ($p$) wave mode conversion
coefficient $A_1$ ($A_2$) as a function of the parameter $v$, which
is proportional to $\sin\theta$, for several values of $w$ and for
$\zeta=500$, $\tilde{\nu}=10^{-8}$ and $L/\Lambda=20$. In the
absence of the external magnetic field, the $s$ wave mode conversion
coefficient $A_1$ vanishes, whereas the $p$ wave mode conversion
coefficient $A_2$ agrees with the result for unmagnetized cases
reported in Ref.~7. When $B_0>0$, $A_1$ becomes nonzero. For $w$
values well over 1, mode conversion occurs in a narrow range of $v$
or $\theta$. This range is called the radio window and becomes
narrower as $w$ or $\zeta$ increases. In Fig.~5, we plot the
reflectances $R_{11}$, $R_{22}$ and $R_{12}$ for $\zeta=500$,
$\tilde{\nu}=10^{-8}$ and $L/\Lambda=20$. The parameter $w$ is equal
to 0.5 in Fig.~5(a), 1 in Fig.~5(b) and 2 in Fig.~5(c).

The occurrence of radio windows is closely related to the coupling
between the O mode and the X mode near the position where
$\omega=\omega_p$, which has the coordinate $z= L-\Lambda$ ($\equiv
z_0$) in our profile. It is well-known that this coupling is most
efficient when the incident angle satisfies
\begin{equation}
\sin\theta=\pm \sqrt{\frac{Y}{Y+1}}\sin\alpha \label{eq:win1}
\end{equation}
for $Y<1$, or
\begin{equation}
\sin\theta=\pm \sqrt{\frac{Y}{Y-1}}\sin\alpha \label{eq:win2}
\end{equation}
for $Y>1$, where $\alpha$ is the angle between the external magnetic
field and the direction of inhomogeneity and is equal to $90^\circ$
in our case.\cite{budden,ginz,mjol,window1,mjol2,seliga} Then it is
easy to see that there is no incident angle satisfying
Eq.~(\ref{eq:win2}) and that Eq.~(\ref{eq:win1}) can be rewritten as
\begin{equation}
v=\frac{w}{\sqrt{1+w^2/\zeta^{2/3}}}, \label{eq:win3}
\end{equation}
where we have kept only the plus sign.

In Fig.~\ref{f.6}(a), we plot $v_{1,{\rm max}}$ and $v_{2,{\rm
max}}$, which are the $v$ values where $A_1$ and $A_2$ take the
maximum values $A_{1,{\rm max}}$ and $A_{2,{\rm max}}$ respectively
in Fig.~\ref{f.4}, versus $w$ for $\zeta=500$, $\tilde{\nu}=10^{-8}$
and $L/\Lambda=20$ and compare them with Eq.~(\ref{eq:win3}). We
find that when $w$ is greater than about 1, the agreement between
our result and Eq.~(\ref{eq:win3}) is excellent. When $w <1$,
however, our result differs greatly from Eq.~(\ref{eq:win3}). In
Fig.~\ref{f.6}(b), the maximum mode conversion coefficients
$A_{1,{\rm max}}$ and $A_{2,{\rm max}}$ are plotted versus $w$.

In the rest of this subsection, we consider the spatial dependence
of electric and magnetic field intensities inside the inhomogeneous
plasma. Using Eqs.~(\ref{eq:imbedf}) and (\ref{eq:matrix}), we have
calculated the electric and magnetic field components $E_{y1}(z)$
and $B_{y1}(z)$ for incident $s$ waves and $E_{y2}(z)$ and
$B_{y2}(z)$ for incident $p$ waves. In Fig.~\ref{f.7}, we plot the
absolute values of $E_{y1}(z)$, $B_{y1}(z)$, $E_{y2}(z)$ and
$B_{y2}(z)$ for $\zeta=50$, $\tilde{\nu}=10^{-3}$ and
$L/\Lambda=20$. $v$ is chosen to be equal to the critical value
$v_c\equiv w/\sqrt{1+w^2/\zeta^{2/3}}$ and the value of $w$ is 1 in
Figs.~\ref{f.7}(a-d), 2 in Figs.~\ref{f.7}(e-h) and 3 in
Figs.~\ref{f.7}(i-l). On these figures, we also indicate the
positions of the resonant point $z_R$ given by Eq.~(\ref{eq:respt})
and the point $z_c$ where incident waves start to become evanescent
when Eq.~(\ref{eq:win3}) is satisfied. $z_c$ turns out to be given
by
\begin{equation}
z_c -z_0= \Lambda\frac{2Y}{1+Y}, \label{eq:zc}
\end{equation}
where $Y=w^2/\zeta^{2/3}$. For $z_R<z<z_c$, electromagnetic waves
are expected to be evanescent. The highly oscillating behavior for
$z<z_R$ in Figs.~\ref{f.7}(i-l) comes from a branch of the X mode
commonly known as the Z mode.

In Fig.~\ref{f.8}, we compare the field distributions at the
critical value of $v$ given by Eq.~(\ref{eq:win3}) with those for
$v$ slightly different from the critical value. In
Figs.~\ref{f.8}(e-h), $\zeta=50$, $\tilde{\nu}=10^{-3}$,
$L/\Lambda=20$, $w=3$ and $v=v_c\approx 2.326$. $v$ is equal to
$v_c-0.1$ in Figs.~\ref{f.8}(a-d) and $v_c+0.1$ in
Figs.~\ref{f.8}(i-l) and all other parameters are the same as in
Figs.~\ref{f.8}(e-h). We find that the evanescent behavior is absent
in off-critical cases.

In Fig.~\ref{f.9}, we show the absolute values of the $z$ components
of the electric field, $E_{z1}$ for incident $s$ waves and $E_{z2}$
for incident $p$ waves, calculated using Eq.~(\ref{eq:xz}). The
parameters used are $\zeta=50$, $\tilde{\nu}=10^{-3}$,
$L/\Lambda=20$, $w=3$ and $v=v_c$ for Figs.~\ref{f.9}(c-d),
$v=v_c-0.3$ for Figs.~\ref{f.9}(a-b), $v=v_c+0.3$ for
Figs.~\ref{f.9}(e-f). In off-critical cases, the evanescent behavior
is absent and the divergence at the resonance point becomes weaker.

\subsection{Frequency dependence}\label{sec4c}

For the study of the frequency dependence of mode conversion
phenomena, we use the plasma density profile given by
Eq.~(\ref{eq:profile2}). We assume $f_0=1$ MHz, so the local plasma
frequency varies from 1 MHz at $z=L$ to 5 MHz at $z=0$. The
cyclotron frequency $f_c$ is 1.5 MHz and the collision frequency
$f_\nu$ is equal to $10^{-8}$ MHz. We show the mode conversion
coefficients $A_1$ and $A_2$ for several incident angles and for
$L=250$ m in Fig.~\ref{f.10} and for $L=1500$ m in Fig.~\ref{f.11}.
In all cases, mode conversion is found to occur only in the
frequency range determined by the upper hybrid frequency. In other
words, $A_1$ and $A_2$ are nonzero only for $\sqrt{1^2+1.5^2}\approx
1.8~ {\rm MHz} <f<\sqrt{5^2+1.5^2}\approx 5.22~ {\rm MHz}$.

In the case of uniform magnetized plasmas, the O (X) wave shows a
resonance when the wave frequency $\omega$ is equal to the resonance
frequency $\omega_O$ ($\omega_X$). The frequencies $\omega_O$ and
$\omega_X$ are given by
\begin{widetext}
\begin{eqnarray}
&&\omega_O\equiv \sqrt{\frac{\omega_p^2+\omega_c^2}{2}
-\sqrt{\left(\frac{\omega_p^2+\omega_c^2}{2}\right)^2
-\omega_p^2\omega_c^2\cos^2\phi}},\nonumber\\
&&\omega_X\equiv \sqrt{\frac{\omega_p^2+\omega_c^2}{2}
+\sqrt{\left(\frac{\omega_p^2+\omega_c^2}{2}\right)^2
-\omega_p^2\omega_c^2\cos^2\phi}},
\end{eqnarray}
\end{widetext}
where $\phi$ is the angle between the external magnetic field and
the wave vector.\cite{budden} The fact that mode conversion occurs
only when $\omega$ is equal to $\omega_{uh}$ regardless of the
incident angle suggests that in our stratified plasma, only the X
wave component propagating in the direction of inhomogeneity with
$\phi=90^\circ$ can cause mode conversion.

\section{Conclusion}\label{sec5}

In this paper, we have presented a new version of the invariant
imbedding theory for the propagation of coupled waves in
inhomogeneous media and applied it to the mode conversion of high
frequency electromagnetic waves into electrostatic modes in cold,
magnetized and stratified plasmas. We have considered the cases
where the external magnetic field is applied perpendicularly to the
direction of inhomogeneity and the electron density profile is
linear. We have obtained extensive and numerically exact results for
the mode conversion coefficients, the reflectances and the wave
electric and magnetic field profiles inside the inhomogeneous
plasma. We have explored the dependences of mode conversion
phenomena on the size of the external magnetic field, the incident
angle and the wave frequency in detail. Our theoretical method and
results are expected to be quite useful in investigating a broad
range of observations and experiments on the interaction between
electromagnetic waves and plasmas both in space and laboratories. In
forthcoming papers, we will apply our theory to the cases where the
external magnetic field is applied in different directions and to
the propagation and mode conversion of low frequency waves in more
general plasmas consisting of various kinds of ions as well as
electrons.

\acknowledgments This work has been supported by the ABRL program
through grant number R14-2002-062-01000-0.

\appendix
\section{Derivation of the matrices $\mathcal{E}$ and $\mathcal{M}$}
\label{sec6}

Using the Maxwell's equations
\begin{eqnarray}
&&\nabla\times{\bf E}=-{1 \over c}{{\partial{\bf B}}\over{\partial
t}}=ik{\bf B},\nonumber\\&&\nabla\times{\bf B}={1 \over
c}{{\partial{\bf D}}\over{\partial t}}=-ik\epsilon\cdot{\bf E}
\end{eqnarray}
and the dielectric tensor, Eq.~(\ref{eq:dtensor}), we express the
$z$-dependent field amplitudes $E_x$, $E_z$, $B_x$ and $B_z$ in
terms of $E_y$ and $B_y$:
\begin{eqnarray}
&&E_x(z)=-\frac{i}{k\epsilon_3(z)}{B_y}^\prime(z),\nonumber\\
&&E_z(z)=i\frac{\epsilon_2(z)}{\epsilon_1(z)}E_y(z)
-\frac{\sin\theta}{\epsilon_1(z)}B_y(z),\nonumber\\
&&B_x(z)=\frac{i}{k}{E_y}^\prime(z),\nonumber\\
&&B_z(z)=E_y(z)\sin\theta, \label{eq:xz}
\end{eqnarray}
where the prime denotes a differentiation with respect to $z$. We
substitute these equations and their derivatives into the $z$
components of Eq.~(\ref{eq:cwe}) and obtain a matrix equation of the
form
\begin{eqnarray}
\psi^{\prime\prime}+{\mathcal A}\psi^\prime+{\mathcal B}\psi=0,
\end{eqnarray}
where
\begin{widetext}
\begin{eqnarray}
&&\psi=\pmatrix{ E_y(z) \cr B_y(z) \cr},
~~{\mathcal A}=\pmatrix{0&0\cr 0&-\frac{{\epsilon_3}^\prime(z)}{\epsilon_3(z)}\cr},\nonumber\\
&&{\mathcal
B}=k^2\pmatrix{\epsilon_1(z)-\frac{\left[\epsilon_2(z)\right]^2}{\epsilon_1(z)}-\sin^2\theta
& -i{\epsilon_2(z)\over\epsilon_1(z)}\sin\theta \cr
i{\epsilon_2(z)\epsilon_3(z)\over\epsilon_1(z)}\sin\theta
&\epsilon_3(z)\left[1-\frac{\sin^2\theta}{\epsilon_1(z)}\right]\cr}.
\label{eq:ab}
\end{eqnarray}
\end{widetext}
Comparing this with Eq.~(\ref{eq:wave2}), we get
${\mathcal A}=-{\cal E}^\prime{\cal E}^{-1}$ and ${\mathcal B}={\cal
E}K^2{\cal M}=k^2{\cal E}{\cal M}$ with $K=kI$. From the forms of
$\mathcal A$ and $\mathcal B$, we deduce the matrices
\begin{eqnarray}
&&{\cal E}=\pmatrix{1&0\cr 0&\epsilon_3\cr},\nonumber\\
&&{\cal
M}=\pmatrix{\epsilon_1-\frac{{\epsilon_2}^2}{\epsilon_1}-\sin^2\theta
& -i{\epsilon_2\over\epsilon_1}\sin\theta \cr
i{\epsilon_2\over\epsilon_1}\sin\theta &
1-\frac{\sin^2\theta}{\epsilon_1}\cr}
\end{eqnarray}
in a straightforward manner.

\newpage

\begin{figure}
\includegraphics[width=8cm]{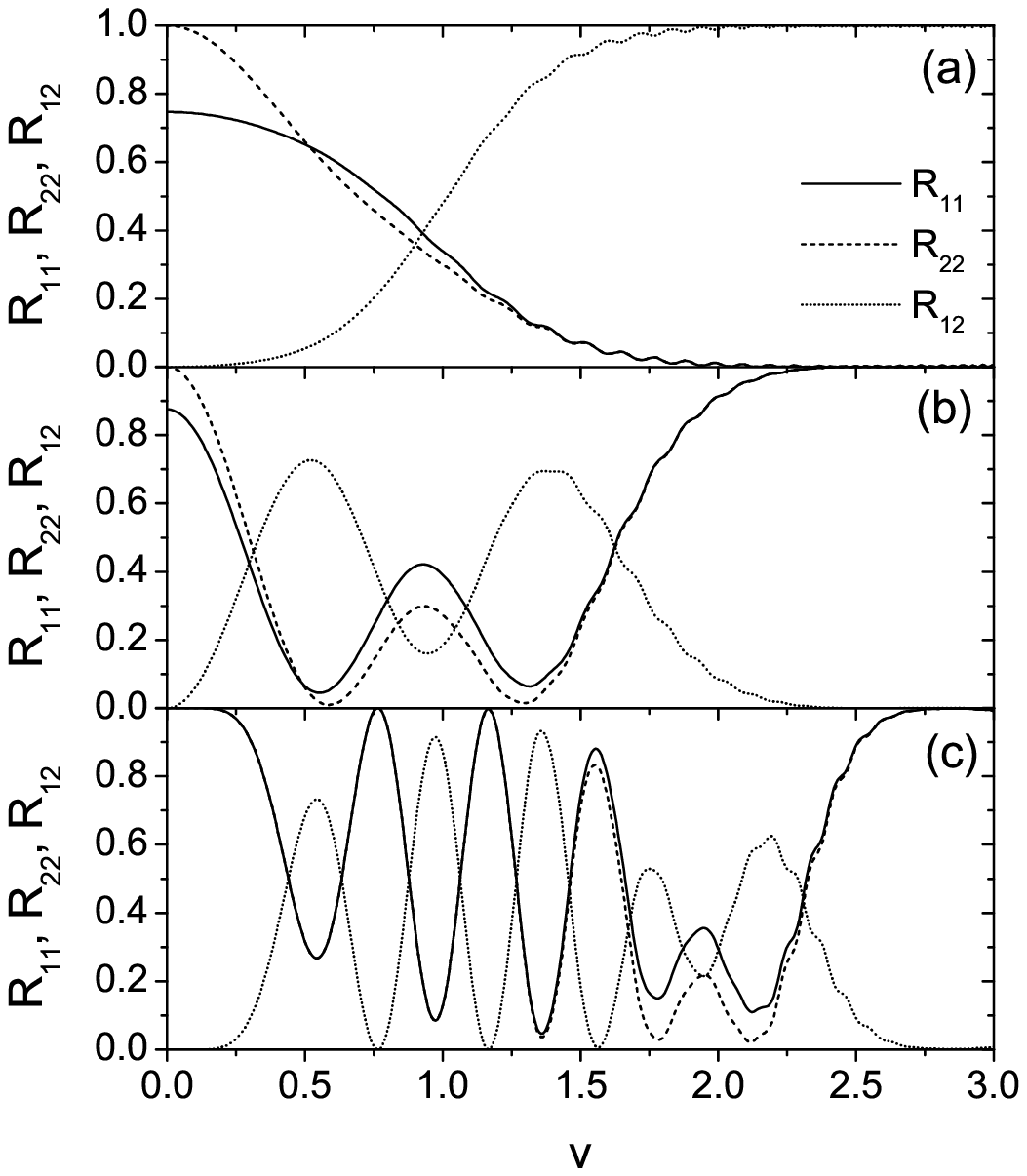}
\caption{Reflectances $R_{11}$, $R_{22}$ and $R_{12}$ ($=R_{21}$) as
a function of $v$ for $\zeta=500$, $\tilde{\nu}=10^{-8}$,
$L/\Lambda=20$ and (a) $w=0.5$, (b) $w=1$, (c) $w=2$.} \label{f.5}
\end{figure}

\begin{figure}
\includegraphics[width=8cm]{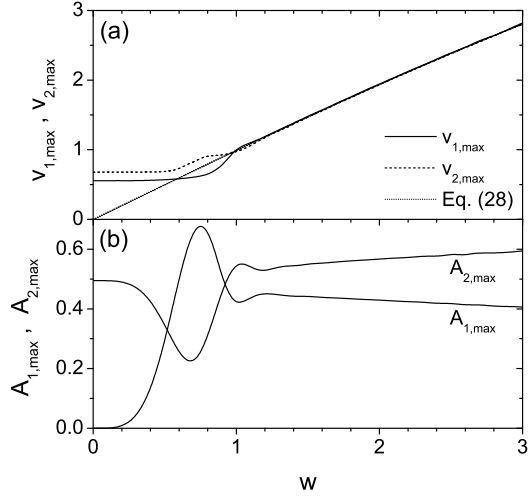}
\caption{(a) $v_{1,{\rm max}}$ ($v_{2,{\rm max}}$), which is the $v$
value where $A_1$ ($A_2$) takes the maximum value $A_{1,{\rm max}}$
($A_{2,{\rm max}}$) in Fig.~\ref{f.4}, as a function of $w$ for
$\zeta=500$, $\tilde{\nu}=10^{-8}$ and $L/\Lambda=20$. Our exact
results are compared with Eq.~(\ref{eq:win3}). (b) $A_{1,{\rm max}}$
and $A_{2,{\rm max}}$ versus $w$.}\label{f.6}
\end{figure}

\begin{figure}
\includegraphics[width=8cm]{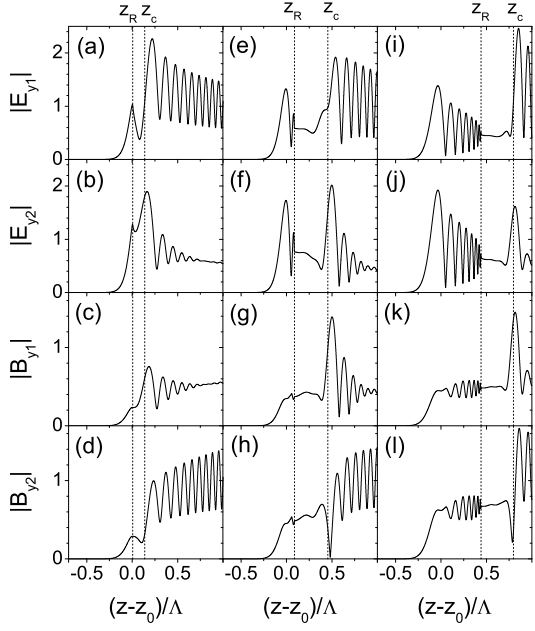}
\caption{Spatial dependence of the absolute values of the electric
and magnetic field components $E_{y1}(z)$ and $B_{y1}(z)$ for
incident $s$ waves and $E_{y2}(z)$ and $B_{y2}(z)$ for incident $p$
waves, for $\zeta=50$, $\tilde{\nu}=10^{-3}$, $L/\Lambda=20$,
$v=v_c\equiv w/\sqrt{1+w^2/\zeta^{2/3}}$ and (a-d) $w=1$, (e-h)
$w=2$, (i-l) $w=3$. $z_R$ and $z_c$ are defined by
Eqs.~(\ref{eq:respt}) and (\ref{eq:zc}) respectively. For
$z_R<z<z_c$, the waves are expected to be evanescent.} \label{f.7}
\end{figure}

\begin{figure}
\includegraphics[width=8cm]{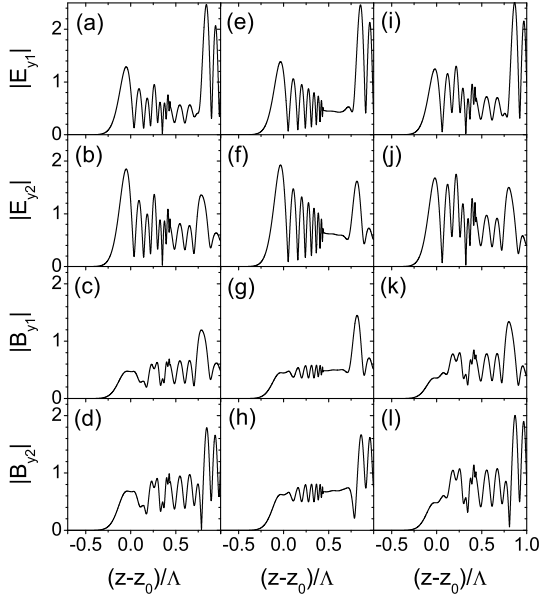}
\caption{Spatial dependence of the absolute values of the electric
and magnetic fields for $\zeta=50$, $\tilde{\nu}=10^{-3}$,
$L/\Lambda=20$, $w=3$ and (a-d) $v=v_c-0.1$, (e-h) $v=v_c$, (i-l)
$v=v_c+0.1$.} \label{f.8}
\end{figure}

\begin{figure}
\includegraphics[width=8cm]{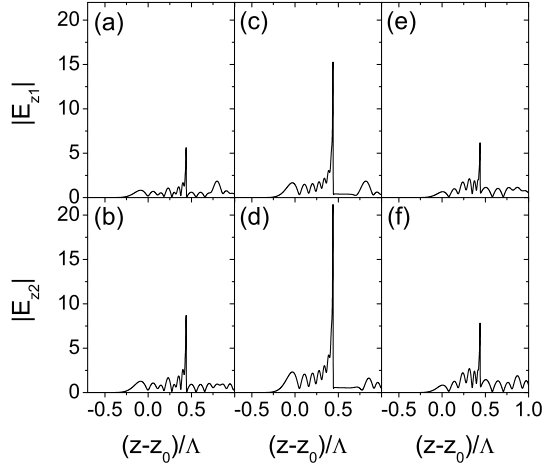}
\caption{Spatial dependence of the absolute values of the $z$
components of the electric field for $\zeta=50$,
$\tilde{\nu}=10^{-3}$, $L/\Lambda=20$, $w=3$ and (a-b) $v=v_c-0.3$,
(c-d) $v=v_c$, (e-f) $v=v_c+0.3$.} \label{f.9}
\end{figure}

\begin{figure}
\includegraphics[width=8cm]{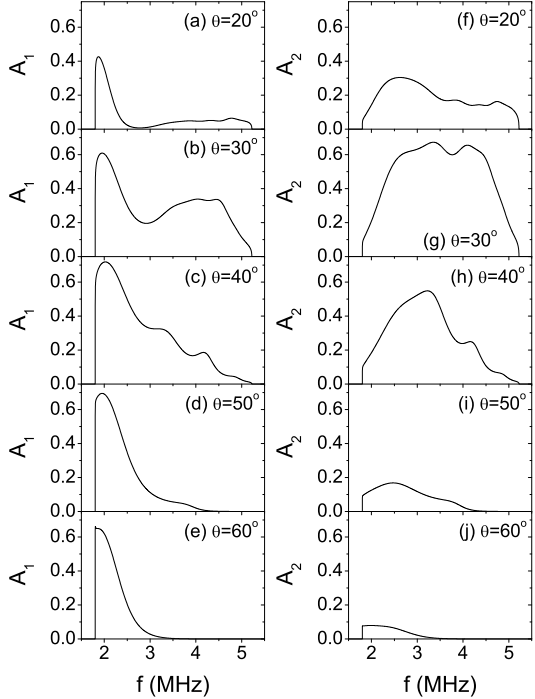}
\caption{Frequency dependence of the mode conversion coefficients
$A_1$ and $A_2$ for the linear profile given by
Eq.~(\ref{eq:profile2}) and for incident angles $\theta=20^\circ$,
$30^\circ$, $40^\circ$, $50^\circ$ and $60^\circ$. $f_0=1$ MHZ,
$f_c=1.5$ MHZ, $f_\nu=10^{-8}$ MHz and $L=250$ m. Note that mode
conversion is limited to the frequency range given by the upper
hybrid frequency, that is $\sqrt{1^2+1.5^2}\approx 1.8~ {\rm MHz}
<f<\sqrt{5^2+1.5^2}\approx 5.22~ {\rm MHz}$.} \label{f.10}
\end{figure}

\begin{figure}
\includegraphics[width=8cm]{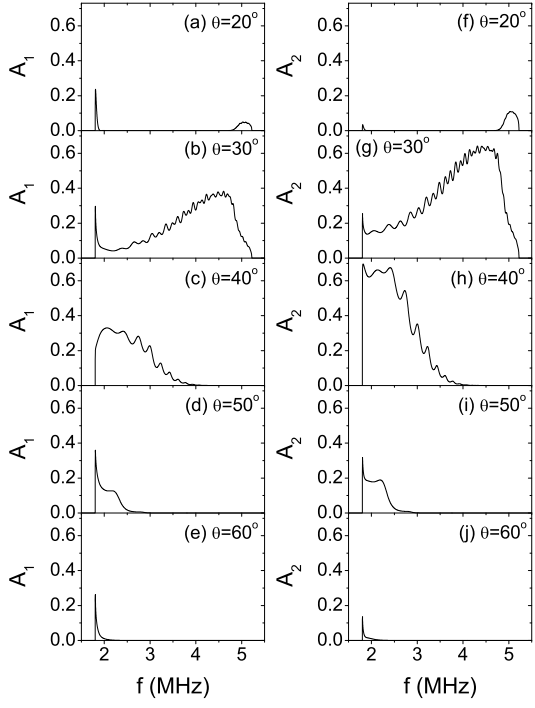}
\caption{Frequency dependence of the mode conversion coefficients
$A_1$ and $A_2$ for the linear profile given by
Eq.~(\ref{eq:profile2}) and for incident angles $\theta=20^\circ$,
$30^\circ$, $40^\circ$, $50^\circ$ and $60^\circ$. $f_0=1$ MHZ,
$f_c=1.5$ MHZ, $f_\nu=10^{-8}$ MHz and $L=1500$ m.} \label{f.11}
\end{figure}

\end{document}